# *In silico* ADMET and molecular docking study on searching potential inhibitors from limonoids and triterpenoids for COVID-19

Seshu Vardhan and Suban K Sahoo[*]

*Department of Applied Chemistry, S.V. National Institute of Technology (SVNIT), Surat-395007, India. (E-mail: sks@chem.svnit.ac.in; subansahoo@gmail.com)*

**Abstract**

Virtual screening of phytochemicals was performed through molecular docking, simulation, *in silico* ADMET and drug-likeness prediction to identify the potential hits that can inhibit the effects of SARS-CoV-2. Considering the published literature on medicinal importance, total 154 phytochemicals with analogous structure from limonoids and triterpenoids were selected to search potential inhibitors for the five therapeutic protein targets of SARS-CoV-2, i.e., 3CLpro (main protease), PLpro (papain-like protease), SGp-RBD (spike glycoprotein-receptor binding domain), RdRp (RNA dependent RNA polymerase) and ACE2 (angiotensin-converting enzyme 2). The *in silico* computational results revealed that the phytochemicals such as glycyrrhizic acid, limonin, 7-deacetyl-7-benzoylgedunin, maslinic acid, corosolic acid, obacunone and ursolic acid were found to be effective against the target proteins of SARS-CoV-2. The protein-ligand interaction study revealed that these phytochemicals bind with the amino acid residues at the active site of the target proteins. Therefore, the core structure of these potential hits can be used for further lead optimization to design drugs for SARS-CoV-2. Also, the medicinal plants containing these phytochemicals like licorice, neem, tulsi, citrus and olives can be used to formulate suitable therapeutic approaches in traditional medicines.

# 1. Introduction

There is extensive ongoing research globally to formulate suitable therapeutic approaches to control the effects of the severe acute respiratory syndrome coronavirus 2 (SARS-CoV-2) to human life that caused the disease COVID-19. The first patient infected with SARS-CoV-2 was detected in December, 2019 at Wuhan, China [1]. Subsequently, the virus spread across 187 countries and territories due to its high human to human contagious nature and infected 10710005 as of 3rd July 2020 with a total death of 517877 [2]. The World Health Organization (WHO) declared the outbreak of SARS-CoV-2 as a public health emergency of international concern (PHEIC) on 30th Jan 2020, and a pandemic on 11th March 2020 [3]. The non-availability of medically proven efficacious drugs or vaccines is the main concern of COVID-19 pandemic [4]. Therefore, effective steps to identify the infected persons through rapid diagnosis followed by quarantined them to stop the further spread of the virus are underway to fight against this pandemic. In addition, the FDA-approved drugs like hydroxychloroquine, remdesivir, favipiravir, arbidol, hydroxychloroquine/azithromycin, lopinavir/ritonavir and lopinavir/ritonavir combined with interferon beta etc. are repurposed, but despite some promising results further clinical studies are required to examine their mechanisms of inhibition, efficacy and safety in the treatment of COVID-19 [5-9]. Therefore, both the computational and experimental approaches are adopted to search suitable drugs from the library of FDA-approved drugs and also drugs under clinical trial but not yet repurposed against COVID-19. Simultaneously, some recent reports supporting the use of traditional medicines as an adjuvant for the treatment of COVID-19 [10-12] and therefore, efforts are also going to integrate the use of both western drugs and traditional medicines for formulating suitable therapeutic strategies.

The computational approaches like molecular dynamic simulations, molecular docking, drugs-likeness prediction, *in silico* ADMET study etc. are adopted mainly to screen potential drugs/molecules from various databases/libraries. The computational screening saves the experimental cost and time in the field of drug discovery. Considering the recent results of the use of traditional medicines in managing the COVID-19 epidemic [10-12], the current research work was carried out to screen phytochemicals found mainly in the Indian medicinal plants with the important objectives: (i) to search phytochemicals that bind effectively at the active sites of the therapeutic protein targets of SARS-CoV-2, (ii) to propose important hits that can be further investigated for lead optimization and drug discovery, and (iii) to provide computational evidences for formulating traditional medicines against SARS-CoV-2.

Our literature survey revealed that the triterpenoids like 3β-friedelanol from *Euphorbia neriifolia,* quinone-methide triterpenoids extracted from *Tripterygium regelii* (Celastraceae) and glycyrrhizin from *Glycyrrhiza glabra* are experimentally proven to inhibit the effects of SARS-CoV (first identified in Guangdong, China in 2002) [13-16]. Also, our recent molecular docking studies of phytochemicals against the therapeutic protein targets of SARS-CoV-2 supported the effective binding affinity towards limonin, a triterpenoid found in citrus [17]. The highest level of genomic similarity between SARS-CoV and SARS-CoV-2 [18], and the effectiveness of triterpenoids against SARS-CoV prompted us to search potential phytochemicals from limonoids and triterpenoids. In this manuscript, total 154 phytochemicals from limonoids and triterpenoids were selected by considering their known medicinal importance to search potential hits for the five therapeutic protein targets of SARS-CoV-2 i.e., 3CLpro (main protease), PLpro (papain-like protease), SGp-RBD (spike glycoprotein-receptor binding domain), RdRp (RNA dependent RNA polymerase) and ACE2 (angiotensin-converting enzyme 2). The phytochemicals were screened through *in silico* molecular docking, simulation, ADMET and drugs-likeness prediction to propose the potential hits against SARS-CoV-2.

## 2. Experimental

### 2.1. Compounds and proteins selection

The 154 biologically important phytochemicals from limonoids and triterpenoids were first selected based on their reported medicinal properties. The structures of the phytochemicals were collected from various sources and screened to filter the potential compounds that can inhibit the effects of SARS-CoV-2. The SDF files of the selected phytochemicals were retrieved from EMBL-EBI (www.ebi.ac.uk/chebi/advancedSearchFT.do) and PUBCHEM (https://pubchem.ncbi.nlm.nih.gov/). The collected structures of the phytochemicals were further optimized by semi-empirical PM6 method coded in the computational program Gaussian 09W [19]. The optimized structures were converted to the PDB file format by using the program GaussView 5.0. The crystallography structures of the SARS-CoV-2 protein targets (3CLpro, PDB ID: 6LU7; PLpro, PDB ID: 4MM3; RdRp, PDB ID: 6M71; SGp-RDB, PDB ID: 2GHV; ACE2, PDB ID: 6M17) were retrieved from the PDB database (www.rcsb.org).

### 2.2. Molecular docking and simulations

The molecular docking studies were carried out to estimate the binding energies of the phytochemicals towards the therapeutic protein targets of SARS-CoV-2 by using Autodock Vina 1.1.2 [20]. The proteins 3D structures retrieved from RCSB PDB databases were modelled using Swiss-model online server to generate the fine structures. The missing amino acid residues (51-68, 102-110, 122-127, 895-904) were found in the crystal structure of the RdRp protein (PDB ID: 6M71). The refined protein structures were analysed by using the Ramachandran plot (**Fig. S1-S5**). The PDB files of the phytochemicals and proteins were converted into PDBQT files by using AutoDock tools. The grid box dimensions and the grid map coordinates centre for the random and site specific docking for each protein were summarized in **Table S1**. All docking were performed with Lamarckian genetic algorithm (LGA). The docked structures were analysed by using the BIOVIA Discovery studio visualization tool.

The protein structure flexibility and dynamics simulations were performed using the CABSflex2.0 online simulation tool with the default options [21]. The simulated model generated through trajectory clustering k-medoid method. This tool calculates the protein dynamic simulations at 10 ns, predicts fluctuations and protein aggregation propensity. The root-mean-square fluctuation (RMSF) is generated based on the MD trajectory or NMR ensemble. The RMSF of a residue fluctuation profile can be calculated with the following formula:

$$\text{RMSF} = \sqrt{\frac{1}{N} \sum_{J}^{N} (x_i(j) - <x_i>)^2}$$

Where, $x$ is the residue position (Cα atom) $i$ in the trajectory or NMR ensemble model $j$ and $<>$ denotes the average over the whole trajectory or NMR ensemble. In CABS-flex, the statistical errors of RMSF values are reflected in root mean squared deviations (RMSD) between RMSF profile data. CABS-flex tool detects the unusual dynamic behaviour of the secondary structure of the protein, the higher RMSF or fluctuations during the simulation indicates the greater flexibility.

**2.3. ADMET and drug-likeness prediction**

After the molecular docking studies of 154 phytochemicals with the five protein targets of SARS-CoV-2, the absorption, distribution, metabolism, elimination and toxicity (ADMET) of the 47 best dock scored phytochemicals were screened using the online tool 'http://biosig.unimelb.edu.au/pkcsm/prediction' to predict their important pharmacokinetic properties. ADMET properties include absorption: Caco-2 permeability, water solubility, human intestinal absorption, P-glycoprotein substrate, P-glycoprotein I and II inhibitors, skin permeability; distribution: VDss (volume of distribution), fraction unbound, blood brain barrier (BBB) permeability, CNS permeability; metabolism: cytochrome P450 inhibitors, CYP2D6/CYP3A4 substrate; excretion: renal OCT2 substrate, drug total clearance; toxicity: Rat LD50, AMES toxicity, T. pyriformis toxicity, minnow toxicity, maximum tolerated dose, oral rat chronic toxicity, hepatotoxicity, skin sensitization, hERG I and II inhibitors [22].

Drug-likeness properties were predicted using the online tool molinspiration (https://www.molinspiration.com/cgi-bin/properties). Providing SMILES inputs of selected compounds into the tool calculated the LogP values to predict the molecular properties and bioavailability scores resulting in the drug-likeness of the compound. Based on drug-likeness and bioavailability capabilities, the potential compounds were finalized for further protein-ligand interaction study. The calculations of LogP are based on the formula satisfying lipophilicity, hydrophobicity and polarity of the compound explains the ability of molecules that could bind to hydrophobic sites of target proteins [23].

Lipophilicity = Hydrophobicity – Polarity

$LogP = aV + \Lambda$  (V = Molecular volume, $\Lambda$ = Polarity term)

## 3. Results and discussion

### 3.1. Selection of the phytochemicals

Our recent molecular docking study on searching inhibitors for COVID-19 revealed that the phytochemicals limonin known for inhibiting the replication of retroviruses like HTLV-I and HIV-1 showed the higher dock score towards the three protein targets (spike glycoprotein, RdRp and ACE2) of SARS-CoV-2, and comparatively higher than the drug hydroxychloroquine [17]. Limonin, the highly oxygenated triterpenoid dilactone is the first isolated limonoids and till date more than 300 limonoids are isolated and characterized. The limonoids including the structural analogous triterpenoids found in medicinal plants, such as citrus, neem, basil and licorice etc., are reported for various pharmaceutical properties like antiviral, antifungal, antibacterial, anticancer and antimalarial, and also used routinely in the Indian traditional medicine (ayurveda) to treat various health problems [24]. The structure of important limonoids and triterpenoids were first collected from the databases, i.e., EMBL-EBI and PUBCHEM, then a basic preliminary screening of the phytochemicals were carried out based on the published medicinal importance. Total 154 phytochemicals were selected, and then studies against the five therapeutic protein targets (3CLpro, PLpro, SGp-RBD, RdRp and ACE2) of SARS-CoV-2 (**Scheme 1**).

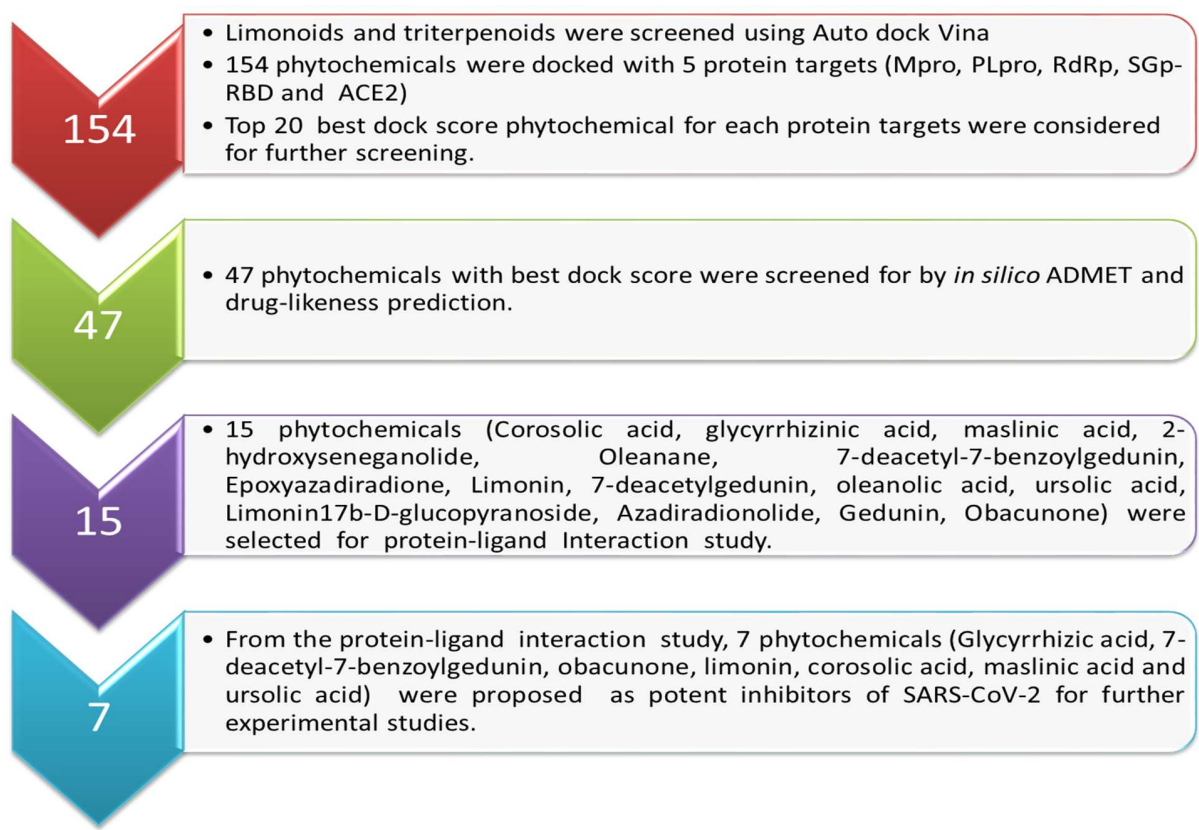

**Scheme 1.** Flowchart showing the steps to screen phytochemicals for the COVID-19.

### 3.2. Molecular docking results

The selected 154 phytochemicals were screened against the five important protein targets, i.e., Mpro or 3CLpro, PLpro, SGp-RBD, RdRp and ACE2 of SARS-CoV-2 by performing random molecular docking using the Autodock Vina computational program. The study of the mechanism of action of the SARS-CoV-2 virus for infecting human cells revealed that the structural spike glycoprotein (S protein) of SARS-CoV-2 interacts with the transmembrane protein of host cell receptor ACE2 [25,26]. This process also internalizes the virus into the endosomes, where the conformational changes take place in the spike glycoprotein that allowed the virus to enter into the human host cell. Thereafter, the RdRp facilitates the viral genome replication [27]. The 3CLpro and PLpro act as proteases in the process of proteolysis of the viral polyprotein into functional units [28]. In short, the SGp and ACE2 are collectively involved in disease establishment and the 3CLpro, PLpro, RdRp involved in translation and replication lead to virus proliferation in the host cell. Therefore, these five proteins were considered as the therapeutic protein targets for the molecular docking with the selected 154 phytochemicals.

The dock score of the phytochemicals against each protein is summarized in **Table S2**. The table of dock score of 154 phytochemicals against the five target proteins revealed that majority of the phytochemicals showed dock score higher than -6.5 kcal/mol [29], and comparably higher dock score than the western drugs hydroxychloroquine, remdesivir and arbidol studied as a control [30]. As the core part of the structure of all phytochemicals are similar, the best 20 phytochemicals for each protein that showed higher dock score were selected for further *in silico* ADMET and drug-likeness study (**Table S3**).

### 3.3. *In silico* ADMET and drug-likeness results

Some phytochemicals commonly showed higher dock score with multiple protein targets and therefore, selecting best 20 phytochemicals for each protein target resulted total 47 phytochemicals. The 47 compounds selected based on their higher dock score were screened further for *in silico* ADMET and drug-likeness study. Out of 47, only 15 compounds are obeying the ADMET limitations and drug-likeness LogP values (**Table 1**). These compounds satisfied the limitations of lipophilicity, hydrophobicity and polarity. The drug-likeness properties are screened based on miLogP (molinpiration LogP) values and TPSA (topological polar surface area) [31]. This study helps in screening out the best compound with drug-likeness and polarity of compound permeable in biological system. The results of ADMET properties and BOILED-Egg model of the 15 phytochemicals are summarized in **Table S4** and **Table S5**. ADMET results are interpreted based on marginal values compared with resultant value as high caco-2 permeability predicted value >0.90, intestinal absorption less than 30% is considered as poorly absorbed, human VDss is low if it is below 0.71 L/kg and high value above 2.81L/kg, BBB permeability logBB > 0.3 considered as it cross BBB and logBB<-1 are poorly distributed. CNS permeability interpreted through logPS > -2 penetrate CNS and logPS < -3 as unable to penetrate. T. pyriformis toxicity predicted value > -0.5 ug/L considered as toxic and Minnow toxicity log LC50< -0.3 considered as high acute toxicity [22]. Most of the potent phytochemicals finalised are found in the extracts of medicinal plants like neem, basil, licorice, olives and citrus. These 15 phytochemicals were selected for protein-ligand interaction study to identify the potential hits that bind at the active sites of the respective protein targets of SARS-CoV-2.

**Table 1.** List of phytochemicals screened based on *in silico* ADMET, drug-likeness and published pharmaceutical data.

| Compounds | Sources | Medicinal Properties | miLogP | TPSA | Ref. |
|---|---|---|---|---|---|
| Corosolic acid | Lagerstroemia speciosa | Supress proliferation of cancer cells | 5.87 | 77.75 | 32 |
| Glycyrrhizic acid | Licorice | Treats liver diseases, Anti HIV-1, SARS-CoV | 1.97 | 267.04 | 33,34 |
| Maslinic acid | Olives | Anti-oxident, anti-inflammatory, weak inhibition to cytochrome P450 | 5.81 | 77.75 | 35 |
| 2-Hydroxyseneganolide | Fruits of Khaya senegalensis | Antifungal activity especially against botrytis cinerea | 1.47 | 132.51 | 36 |

| | woody angiosperms | Antioxidant, anti-inflammatory, hepatoprotective, cardioprotective, antipruritic, spasmolytic, antiallergic, antimicrobial and antiviral effects and anti canceral especially breast cancer | 8.86 | 0 | 37 |
|---|---|---|---|---|---|
| Oleanane | | | | | |
| 7-Deacetyl-7-benzoylgedunin | Neem (Azadirachta indica) | Activity against HL60 leukemia cells | 6.07 | 95.35 | 38 |
| Epoxyazadiradione | Neem (Azadirachta indica) | Plasmodium falciparum plasmepsin I inhibitor | 3.66 | 86.11 | 39 |
| Limonin | Citrus Fruits | Inhibit the HIV 1 replication in cellular systems | 2.53 | 104.58 | 40 |
| 7-Deacetylgedunin | Neem (Azadirachta indica) | Antimalarial, anti-inflammatory | 3.64 | 89.27 | 41 |
| Oleanolic acid | Ocimum Sanctum (Basil) | Therapeutic potential for neurodegenerative diseases | 6.72 | 57.53 | 42 |
| Ursolic acid | Ocimum Sanctum (Basil) | Therapeutic potential for neurodegenerative diseases | 6.79 | 57.53 | 43 |
| Limonin glucoside | Citrus Fruits | Inhibit colon adenocarcinoma cell proliferation through apoptosis | -0.29 | 214.96 | 44 |
| Azadiradionolide | Neem (Azadirachta indica) | Apoptosis inducing activity | 2.85 | 86.75 | 45 |
| Gedunin | Neem (Azadirachta indica) | Antiplasmodial | 4.34 | 95.35 | 46 |
| Obacunone | Citrus Fruits | Represses solmonella pathogenity and also inhibits human colon cancer | 3.8 | 95.35 | 47 |

**3.4. Protein-ligand interaction study**

**3.4.1. Screening of inhibitors for main protease**

The main protease 3CLpro is a cysteine protease containing three domains, i.e., domains I (8-101 residues) and II (102-184 residues) with antiparallel β-barrel structure and the domain III (201-303 residues) with five α-helices linked to the domain II by a long loop (185-200 residues) [48]. The catalytic dyad CYS145 and HIS41, and the residue GLU166 involved in protein dimerization, substrate cleaving through catalysis present in the cleft between domains I and II. The enzyme active site consists of 6 subunits (S1-S6), and the active site residues are 140-145 and 163-166 amino acids in the domain II. The protein-ligand interaction study revealed that four phytochemicals (7-deacetyl-7-benzoylgedunin, glycyrrhizic acid, limonin and obacunone) bind with the catalytic dyad of 3CLpro. The dock score of the best pose of the four phytochemicals and their important molecular interactions at the active site of 3CLpro is summarized in Table 2. The 7-deacetyl-7-benzoylgedunin binds with higher dock score (-9.1 kca/mol) at the active site followed by the glycyrrhizic acid (-8.7 kca/mol), limonin (-8.7 kca/mol) and obacunone (-7.5 kca/mol).

The docked structure of 7-deacetyl-7-benzoylgedunin binds firmly at the active site of the 3CLpro long loop region of domain II (**Fig. 1a-b**). The catalytic dyad CYS145 and HIS41 respectively formed Pi-alkyl and Pi-Pi T-shaped interactions. In addition, the higher binding affinity of 7-deacetyl-7-benzoylgedunin is attributed to the multiple non-covalent interactions like hydrogen bond, van der Waals (VDW) with other amino acid residues (GLU166, HIS163, ARG188, ASP187, HIS164, GLY143, SER144, LEU141, ASN142, PHE140, HIS172, LEU167, PRO168, MET165) at the active site of 3CLpro. The protein structural simulation generated RMSF graph showing protein residues fluctuation and aggregation (**Fig. S6**). The fluctuated residues showing the hydrophobic cavities where substrate binding and the catalytic functions took place. For 3CLpro simulation, some of the residues fluctuation impact at chain A residues (5-16, 46-56, 136-151, 165-178, 181-196, 241-260, 271-286), and the fluctuation impact showing ligand interactions at these residues. Further, the site specific docking of 7-deacetyl-7-benzoylgedunin was performed at the active site of 3CLpro and the binding of the three best pose with a dock score of -9.1, -8.0 and -7.7 kcal/mol is shown in **Fig. 1c**. In addition to the 7-deacetyl-7-benzoylgedunin, the phytochemicals glycyrrhizic acid, limonin and obacunone are also binding at the active site of 3CLpro (**Table 2**). The binding pose of the best three phytochemicals at the active site of 3CLpro is collectively shown in **Fig. 1d**.

It is also important to mention that the phytochemicals ursolic acid and oleanolic acid isolated from holy basil leaves are effectively binding at the domain III active site residues with the dock score of -8.9 kcal/mol, which enhanced the catalytic activity of 3CLpro. Both the phytochemicals bind with 3CLpro by forming hydrogen bonds with the residues LYS137, LEU272, and the closest non-covalent interactions with the residues THR199, ARG131 and LEU287.

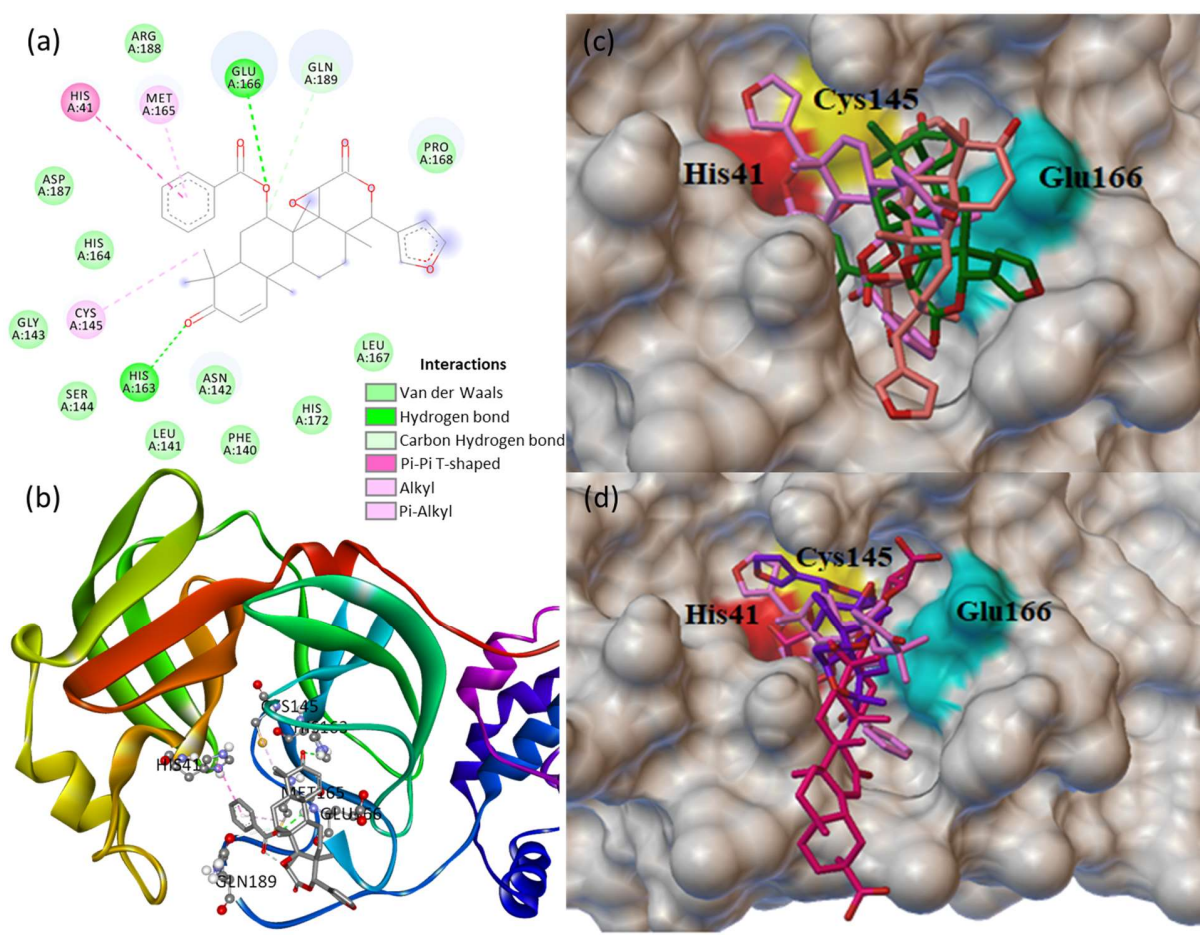

**Fig. 1.** (a) 2D animated pose between 7-deacetyl-7-benzoylgedunin and 3CLpro showing various non-covalent interactions, (b) 3D representation showing the position of 7-deacetyl-7-benzoylgedunin within the hydrophobic cavity of 3CLpro, (c) binding of three best pose of 7-deacetyl-7-benzoylgedunin at the active site of 3CLpro, and (d) binding of 7-deacetyl-7-benzoylgedunin, glycyrrhizic acid and limonin at the active site of 3CLpro.

**Table 2.** The dock score of screened phytochemicals binding at the active site of the main protease 3CLpro and their important interactions with various amino acid residues.

| Phytochemicals | B.E. (kcal/mole) | Important interactions with residues at the active site, catalytic dyed (HIS41 and CYS145) and GLU166 |
|---|---|---|
| 7-deacetyl-7-benzoylgedunin | -9.1 | Carbon hydrogen bond: GLN189; Hydrogen bond: GLU166, HIS163; VDW: ARG188, ASP187, HIS164, GLY143, SER144, LEU141, ASN142, PHE140, HIS172, LEU167; Pi-Pi T-shaped: HIS41; Alkyl: MET165; Pi-Alkyl: CYS145. |
| Glycyrrhizic acid | -8.7 | Hydrogen bond: HIS163, PHE140, GLU166, ASP197; Carbon hydrogen bond: HIS41, GLN189, MET165; VDW: MET49, HIS164, ASP187, ARG187, ARG188, THR190, ALA191, LEU50, HIS172, SER144, LEU141, ASN142. |
| Limonin | -8.7 | Hydrogen bond: GLU166, HIS163, CYS145; Pi-donor: GLY143; Carbon hydrogen bond: GLN189; VDW: ASN142, HIS164, HIS41, MET49. |
| Obacunone | -7.5 | Hydrogen bond: GLU166, HIS163, CYS145; Pi-donor: GLY143; Pi-Alkyl: CYS145; VDW: MET165, GLN189, ASN142, HIS41, HIS164. |

### 3.4.2. Screening of inhibitors for PLpro

PLpro consists of four domains such as thumb, finger, palm and ubiquitin like domain. The active site located in between the thumb and palm domains [49]. The subunits consist of the catalytic triad (CYS112, HIS273 and ASP287), where the active site of PLpro is located. PLpro NSP3 domain contains S2/S4 inhibitor binding sites. Therefore, the molecular screening of phytochemicals that docked at specific residues of S2/S4 site could inhibit the activity of PLpro [49]. The dock score along with the important molecular interactions of the four phytochemicals (obacunone, glycyrrhizic acid, ursolic acid and 7-deacetylgedunin) binding with the catalytic triad at the active site of PLpro are summarized in Table 3. The obacunone (-8.3 kcal/mol) and glycyrrhizic acid (-8.2 kcal/mol) showed almost similar binding affinity followed by ursolic acid (-7.2 kcal/mol) at the pocket of the catalytic triad of PLpro [50,51]. In addition, the protein-ligand interaction study revealed that the phytochemicals epoxyazadiradione and limonin are also binding close to the catalytic site of PLpro.

Obacunone binds firmly at the catalytic site of PLpro, and the docked structure is stabilized by different non-covalent interactions (**Fig. 2a-b**). The catalytic residues HIS273 and ASP287 of PLpro formed Pi-alkyl and carbon hydrogen bond, respectively. Also, the obacunone formed hydrogen bond with residue ARG285, VDW contacts with TRP107, THR266, GLY267, THR275, GLY272, and Pi-alkyl interaction with CYS271. The MD simulations generated the RMSF plot of PLpro showing available contacts to substrate binding at chain B residues (5-10, 170-185, 265-270) (**Fig. S7**). Further, the site specific docking of obacunone was performed and the binding modes of three best pose with dock score -8.3, -7.7 and -7.3 kcal/mol at the catalytic triad of PLpro is shown in **Fig. 2c**. It is also important to mention here that the phytochemicals glycyrrhizic acid, ursolic acid and 7-deacetylgedunin bind with the catalytic residues His273 and Asp287 of PLpro. The binding modes of the best three phytochemicals at the active site of PLpro is collectively shown in **Fig. 2d**.

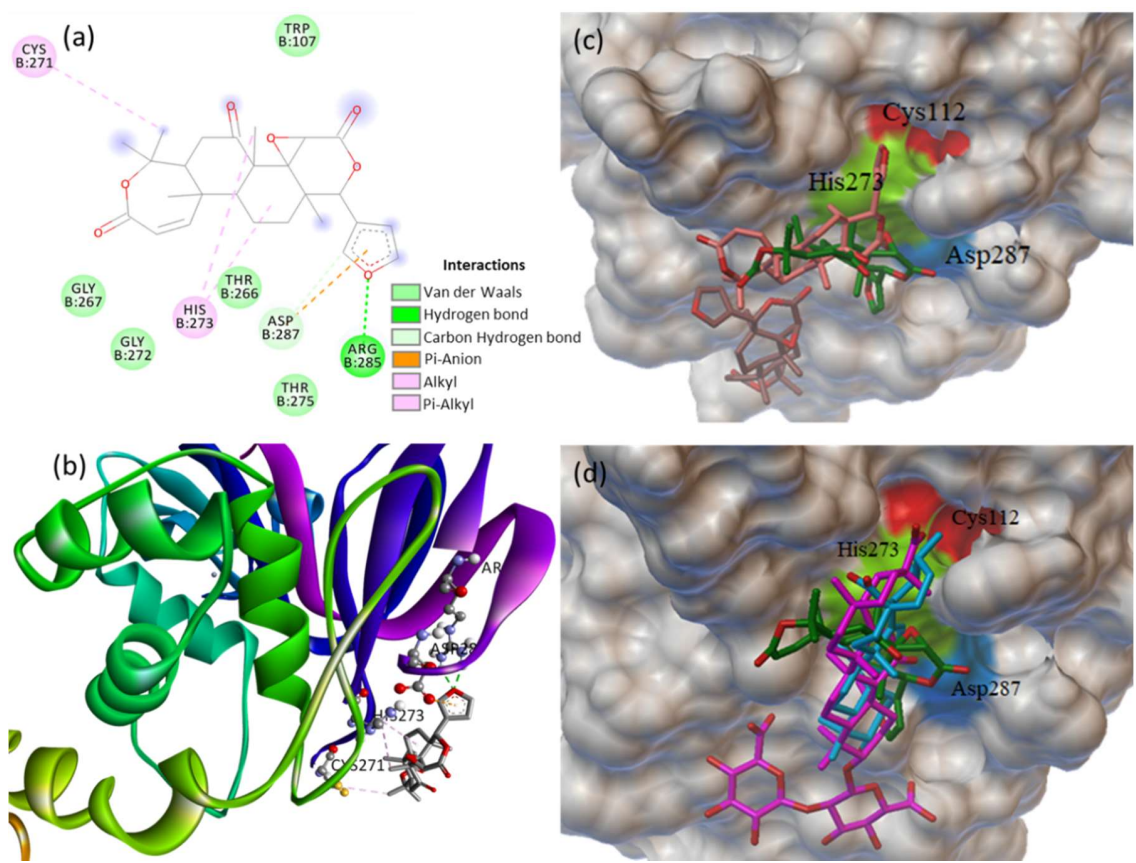

**Fig. 2.** (a) 2D animated pose between obacunone and PLpro showing various non-covalent interactions, (b) 3D representation showing the position of obacunone within the hydrophobic

cavity of PLpro, (c) binding of three best pose of obacunone at the active site of PLpro, and (d) binding of obacunone, glycyrrhizic acid and ursolic acid at the active site of PLpro.

**Table 3.** The dock score of screened phytochemicals binding at the active site of the PLpro and their important interactions with various amino acid residues.

| Phytochemicals | B.E. (kcal/mole) | Important interactions at active site and catalytic triad (CYS112, HIS273, ASP287) |
|---|---|---|
| Obacunone | -8.3 | Hydrogen bond: ARG285; Carbon Hydrogen bond: ASP287; Pi-Alkyl: HIS273, CYS271; Pi-Anion: ASP287, VDW: TRP107, THR266, GLY267, THR275, GLY272. |
| Glycyrrhizic acid | -8.2 | Hydrogen bond: ARG285, TYR297, THR266, Carbon hydrogen bond: LYS298, GLY299, PRO300; Pi-Alkyl: HIS273, CYS271, TRP107; VDW: GLU251, GLU264, MET294, THR292, ASN110. |
| Ursolic acid | -7.2 | Hydrogen bond: ARG285; Pi-Alkyl: HIS273, TRP107; VDW: THR275, ASP287, THR266, CYS271. |
| 7-Deacetylgedunin | -7.1 | Hydrogen bond: THR292, ARG285; Pi-Alkyl: HIS273, Pi-Sigma: HIS290; VDW: LEU291, ASP287, TRP107, THR275, THR266. |

### 3.4.3. Screening of inhibitors for RdRp (NSP12 domain)

RdRp, the non-structural protein NSP12 is a replication tool plays a major role in the transcription cycle of the virus with the help of cofactors NSP7 and NSP8. So, the primary target of the RdRp is NSP12, where the active site is located in between the NiRAN domain β-hairpin [52]. The NTP entry channel is formed by a set of hydrophilic residues such as LYS545, ARG553, and ARG555. The RdRp active site is located in the tunnel-shaped, where the protein complex posing strong electrostatic surfaces contains divalent cationic residues 611-626, especially the residue ASP618. Some of the catalytic residues are located between residues 753-769. More than 80 phytochemicals binds at the RdRp functional sites, but only 6 compounds were finalized based on their high dock score at the active site, *in silico* ADMET and drug-likeness. The interaction details and the binding energy of the 6 phytochemicals at the active site of RdRp are summarized in **Table 4**.

The best phytochemical glycyrrhizic acid is encapsulated in the receptor cavity with the maximum binding energy of -9.9 kcal/mol. The binding site is located between NiRAN domain and β-hairpin structure that polymerizes 3′ end [53], and therefore glycyrrhizic acid may interfere the polymerize activity. Glycyrrhizic acid binds firmly at the active site residues ARG624, ALA762, TRP800, ALA558, SER682, THR556, ARG555, TRP617, GLY616, ASP618, LYS798, VAL763, PHE812, ASP452, GLU811, LYS551, ASP623, VAL557, SER814, SER549, ALA547, ILE548, LYS545 (**Fig. 3a-b**). Glycyrrhizic acid formed non-covalent interaction with divalent cationic residue ASP618, hydrogen bond with ARG555, and VDW with LYS545. Also, the interactions with the catalytic residues, mainly ALA762 makes the glycyrrhizic acid potential phytochemical against RdRp. The three best binding conformations of glycyrrhizic acid at the active site of RdRp is shown in **Fig. 3c**, which clearly indicate that the phytochemical is well inside the hydrophobic cavity created by the residues at the active site of RdRp. Other phytochemicals limonin (-8.2 kcal/mol), 7-deacetyl-7-benzoylgedunin (-8.2 kcal/mol) and limonin glucoside (-8.2 kcal/mol) showed a similar binding affinity at the active site of RdRp. Limonin (furanolactones), the potent phytochemical docked at the active site between NSP12-NSP7 residues formed conventional hydrogen bonds to SER709, LYS714, ASP711, THR710, TRP800 and TRP617, and VDW interactions with residues PRO620, TYR619, ASP760, SER814, ASP761, ASP618, GLY616, GLU811, ALA797 and LYS551, including Pi-alkyl/Pi-Pi contacts with residues CYS622, LYS798 and HIS810 located in RdRp tunnel structure. The collective binding pose of the three best phytochemicals glycyrrhizic acid, limonin and 7-deacetyl-7-benzoylgedunin at the active site of RdRp is shown in **Fig. 3d**. Further, the MD simulations generated the RMSF plot of RdRp showing available contacts to chain A: 31-45, 255-285, 419-459, 909, chain B: 103 and 148-188 residues involved in substrate binding and replication process (**Fig. S8**).

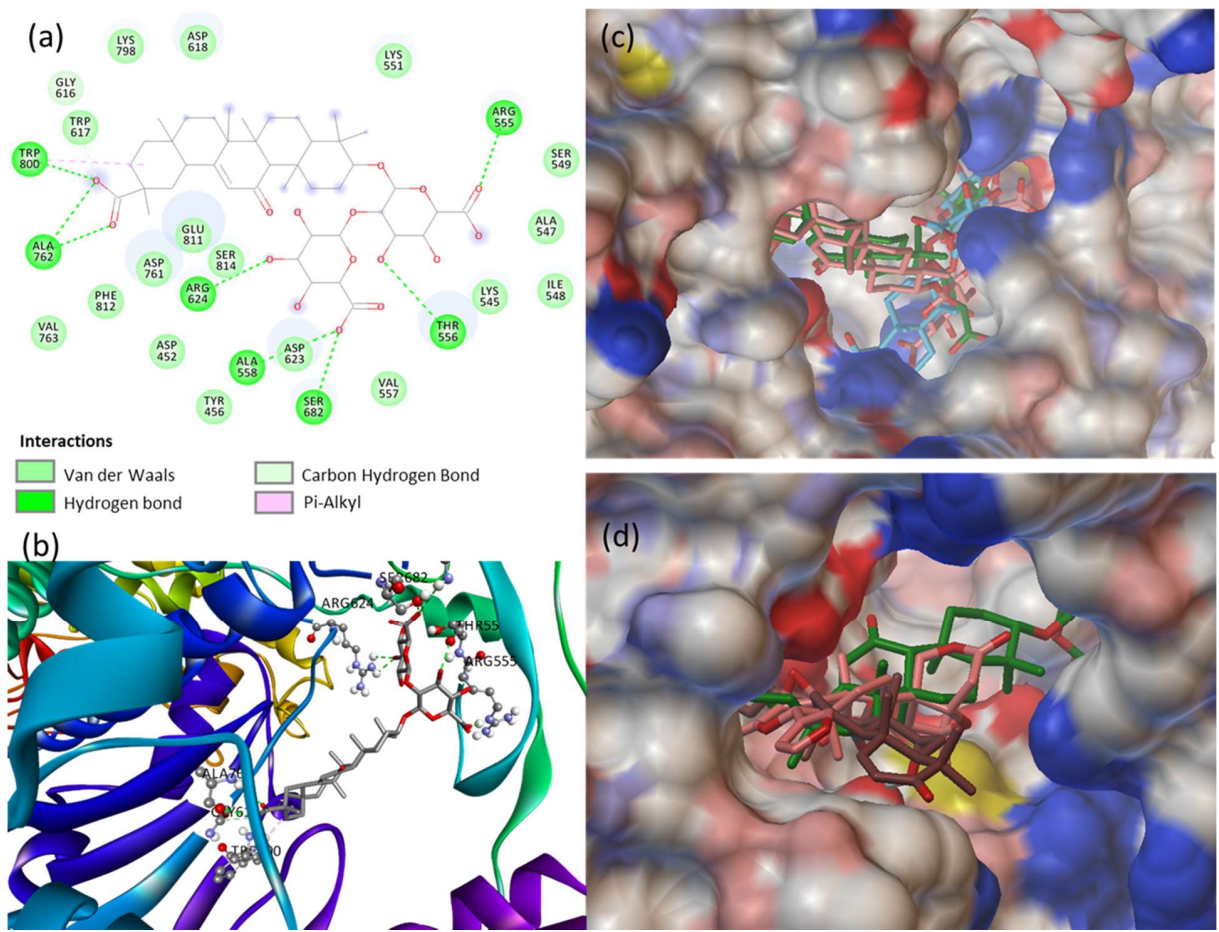

**Fig. 3.** (a) 2D animated pose between glycyrrhizic acid and RdRp showing various non-covalent interactions, (b) 3D representation showing the position of glycyrrhizic acid within the hydrophobic cavity of RdRp, (c) binding of three best pose of glycyrrhizic acid at the active site of RdRp, and (d) binding of glycyrrhizic acid, limonin and 7-deacetyl-7-benzoylgedunin at the active site of RdRp.

**Table 4.** The dock score of screened phytochemicals binding at the active sites of the RdRp and their important interactions with various amino acid residues.

| Phytochemicals | B.E. (kcal/mole) | Important interactions at the active site (residues 611 to 626), divalent cationic residue (ASP618), catalytic site (753-767) and NTP entry channel (LYS545, ARG553, ARG555) |
|---|---|---|
| Glycyrrhizic acid | -9.9 | Hydrogen bond: ARG624, ALA762, TRP800, ALA558, SER682, THR556, ARG555; Carbon hydrogen bond: GLY616; Pi-Alkyl: TRP800; VDW: TRP617, GLY616, ASP618, LYS798, VAL763, PHE812, ASP452, GLU811, LYS551, ASP623, VAL557, SER814, SER549, ALA547, ILE548, LYS545. |
| Limonin | -8.2 | Hydrogen bond: TRP800, TRP617; Pi-Alkyl: CYS622, LYS798; Pi-Pi T-shaped: HIS810; VDW: PRO620, TYR619, ASP760, SER814, ASP761, ASP618, GLY616, GLU811, ALA797, LYS551 |
| 7-deacetyl-7-benzoylgedunin | -8.2 | Hydrogen bond: ALA762; Alkyl/ Pi-Alkyl: LYS798; Carbon hydrogen bond: TRP617, GLU811; Pi-Anion: LYS798, Pi-Cation: ASP761, LYS551, ASP618; VDW: ALA797, TRP800, PHE812, ASP760, TYR619, PRO620. |
| Limonin glucoside | -8.2 | Hydrogen bond: ARG624, ALA554, ARG836; Carbon hydrogen bond: SER549; Alkyl/Pi-Alkyl: HIS439, ALA550, LYS551, ARG555; VDW: ARG553, ASP452, ASP623, TYR456, SER682, VAL557, MET542, LYS545, SER814. |
| 7-deacetylgedunin | -8.1 | Hydrogen bond: ALA762; Carbon hydrogen bond: TRP617, GLU811; Pi-Alkyl: LYS798; Pi-Sigma: TRP800; Pi-Anion: ASP761; VDW: TYR619, ASP760, ASP618, PHE812, ALA797. |
| Obacunone | -7.8 | Hydrogen bond: LYS551, ARG624; Alkyl: ARG555; Pi-Anion: ASP452; VDW: SER549, ARG836, ALA550, ARG553, ALA554, ILE548, THR556, TYR456. |

### 3.4.4. Screening of inhibitors for SGp-RBD

The spike protein determine the virion-host tropism that includes the entry of the virions into the host cells [54]. The receptor binding domain (RBD) of the trimeric spike glycoprotein interacts with the human host cells by binding to ACE2. Due to its structural importance, we focused on SGp-RBD inhibition study by screening 154 phytochemicals through molecular docking, and the binding confirmations were analysed at the active site that could inhibit the SGp-ACE2 complex formation. The RBD of the spike glycoprotein contains 333-527 residues where the active site is located [55]. The protein-ligand interaction study revealed that 6 phytochemicals bind at the active site, and their important molecular interactions are summarized in **Table 5**.

Maslinic acid binds firmly at the active site with a binding energy of -9.3 kcal/mol due to the multiple non-covalent interactions with the residues of SGp-RBD. It forms hydrogen bonds with ASP454, SER456, GLY464, alkyl/Pi-alkyl interactions with HIS445, PHE460, ARG444, PRO477, VAL458, PRO466, and VDW contacts with LEU443, ARG441, LYS465 (**Fig. 4a-b**).

The best three binding poses of masnilic acid at the active site of SGp-RBD with a dock score of -9.3, -8.2 and -7.5 kcal/mol is shown in **Fig. 4c**, which revealed that this phytochemical bind with different residues at the active site of SGp-RBD. MD simulation of SpG-RBD generated the RMSF plot detailing contact sites showing the fluctuations in chain C: 432-438, 352-368, and 456-480 residues of the receptor binding site (**Fig. S9**). In addition to maslinic acid, the phytochemicals glycyrrhizic acid, corosolic acid, 2-hydroxyseneganolide and oleanane showed comparable binding at the active site of SGp-RBD. The binding pose of the best three phytochemicals at the active site of SGp-RBD is shown in **Fig. 4d**.

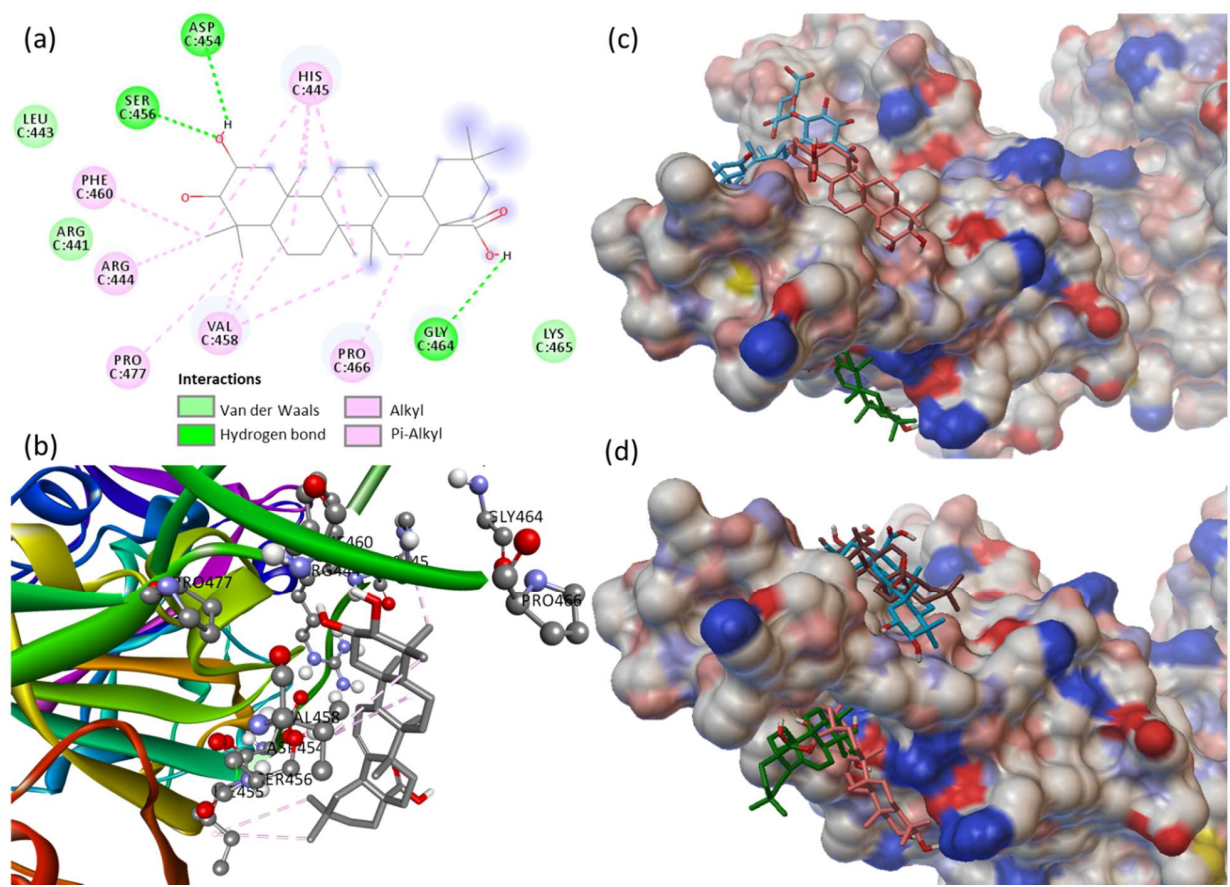

**Fig. 4.** (a) 2D animated pose between maslinic acid and SGp-RBD showing various non-covalent interactions, (b) 3D representation showing the position of maslinic acid within the hydrophobic cavity of SGp-RBD, (c) binding of three best pose of maslinic acid at the active site

of SGp-RBD, and (d) binding of maslinic acid, glycyrrhizic acid and corosolic acid at the active site of SGp-RBD.

**Table 5.** The dock score of screened phytochemicals binding at the active site of the SGp-RBD and their important interactions with various amino acid residues.

| Phytochemicals | B.E. (kcal/mole) | Important interactions at active site of glycosylation (ASN330, TYR356) and ACE2 binding sites (residues 438 to 527) |
|---|---|---|
| Maslinic acid | -9.3 | Hydrogen bond: ASP454, SER456, GLY464; Pi-Alkyl: HIS445, PHE460, ARG444, PRO477, VAL458; Alkyl: PRO466; VDW: LEU443, ARG441, LYS465. |
| Glycyrrhizic acid | -9.3 | Hydrogen bond: PHE360, ARG426, ASN427, TRP423, THR333, SER363; VDW: TYR356, ASN357, SER358, PHE361, SER362, ILE489, GLN492, ASN424, THR359, THR425, ARG495, ILE428, ASN330, ALA332. |
| Corosolic acid | -9.4 | VDW: ASN330, PHE334, ALA331, PHE329, ARG495, TRP423, THR425, THR359, SER358, PHE360, ASN424, ASN427, TYR356, ILE428, THR332. |
| 2-Hydroxyseneganolide | -9.2 | Hydrogen bond: TRP423, VDW: ASN330, PHE329, THR359, THR425, ASN427, ASN424, SER358, TYR356, ASN357, THR332, ALA331, ARG495 |
| Oleanane | -9.0 | VDW: THR332, ALA331, TYR356, ASN330, PHE329, ARG495, TRP423, THR359, ASN424, PHE360, SER358, ASN427, THR425. |
| Gedunin | -8.2 | Hydrogen bond: GLY368; VDW: ASP415, ASP414; Pi-Pi T-shaped: PRO399, LYS365, ALA398. |

### 3.4.5. Screening of inhibitors for ACE2

ACE2 plays a key role in cardio renal disease and acts as a human host receptor for the SARS-CoV-2 [56]. ACE2, a human viral receptor binds to SGp-RBD at a specific site that establishes the primary contact for host-pathogen interaction [56]. The active site residues of ACE2 were studied by using site-directed mutagenesis, and it was found that ARG273 plays a vital role in substrate binding. The HIS345 and HIS505 are catalytic residues plays an important role as a hydrogen bond donor/acceptor to form the tetrahedral peptide intermediate [57]. Also, the residues GLN24, MET82, ILE79, LYS31, HIS34, GLU37, GLY354, GLN325, ASP38, ASN330, GLU329, GLN42 and LEU45 of ACE2 receptor interacts with the SGp-RBD. From the protein-ligand interaction study, the seven phytochemicals bind at the active site of ACE2 and their important molecular interactions are summarized in Table 6. The phytochemicals glycyrrhizic acid, obacunone, azadiradionolide and gedunin bind firmly at the catalytic site of ACE2, whereas maslinic acid, epoxyazadiradione and ursolic acid binds at the RBD site of ACE2.

Glycyrrhizic acid bind firmly at the catalytic site of ACE2 with a dock score of -9.5 kcal/mol. The glycyrrhizic acid interacts to catalytic residues forming hydrogen bonds with ARG273, HIS374, TYR515, ASN394, and VDW contacts with ARG393, TYR385, GLU402, ASP350, ALA348, TRP349, ASP382, HIS505, PHE504 (**Fig. 5a-b**). The three best binding conformations of glycyrrhizic acid at the catalytic site of ACE2 with the dock score of -9.5, -8.3 and -8.2 kcal/mol is shown in **Fig. 5c**. In addition to the glycyrrhizic acid, the phytochemicals obacunone, azadiradionolide and gedunin binds at the catalytic residues ARG273, HIS345 and HIS505 with a binding energy of -8.1, -8.0 and -7.3 kcal/mol, respectively. The binding pose of the three best phytochemicals glycyrrhizic acid, obacunone and azadiradionolide at the catalytic site of ACE2 is shown in **Fig. 5d.** Further, the MD simulation of ACE2 complex with glycyrrhizic acid showed multiple contact sites in receptor chain B: 51-81, 141, 201-231, 261, 321-351, 531-561, 591-651 residues (**Fig. S10**). Out of these residues, the SGp-RBD contacts to ACE2 receptor residues GLN24, MET82, ILE79, LYS31, HIS34, GLU37, GLY354, GLN325, ASP38, ASN330, GLU329, GLN42 and LEU45. It is important to mention here that the phytochemicals maslinic acid, epoxyazadiradione and ursolic acid interact with ACE2 substrate binding site with the dock score of -8.5, -8.0 and -7.4 kcal/mol, respectively. The binding conformation of maslinic acid and epoxyazadiradione at the ACE2 substrate binding site is shown in **Fig. 6**. The effective binding of glycyrrhizic acid, maslinic acid, obacunone, epoxyazadiradione, azadiradionolide, ursolic acid and gedunin at the catalytic site and RBD site of ACE2 may potentially interfere the SGp-RBD-ACE2 complex formation and therefore can prevent the entry of virus into host cells.

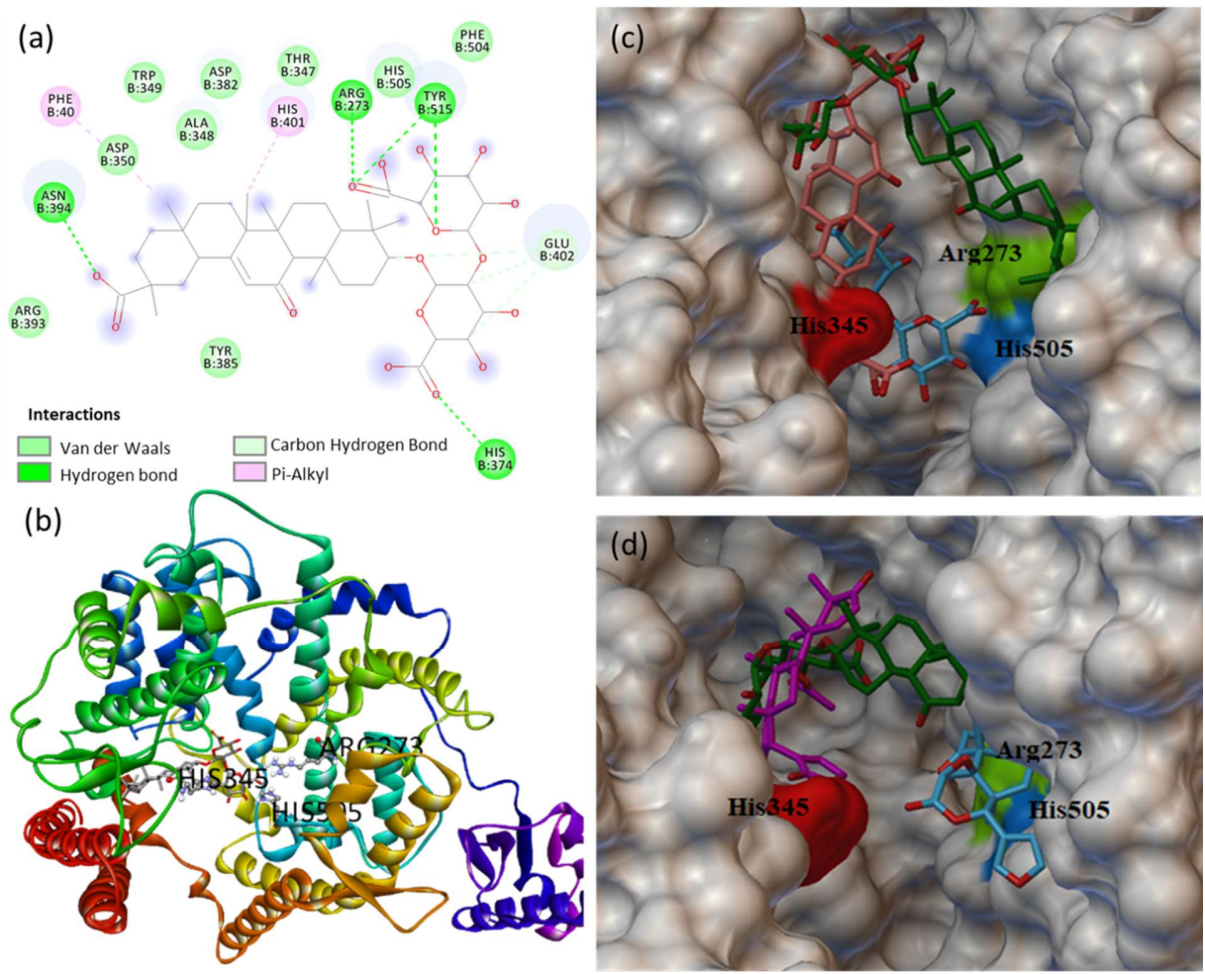

**Fig. 5.** (a) 2D animated pose between glycyrrhizic acid and ACE2 showing various non-covalent interactions at catalytic site and (b) the corresponding 3D representation showing binding conformation. (c) The three best pose of glycyrrhizic acid at the catalytic site of ACE2, and (d) the binding pose of three best phytochemicals glycyrrhizic acid, obacunone and azadiradionolide at the catalytic site of ACE2.

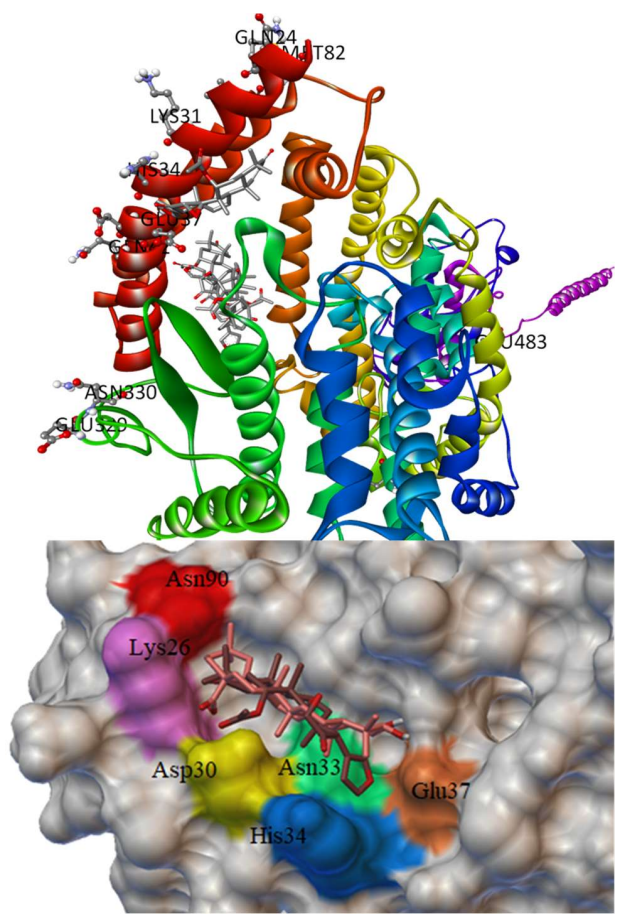

**Fig. 6.** The binding pose of maslinic acid and epoxyazadiradione at the RBD site of ACE2.

**Table 6.** The dock score of screened phytochemicals binding at the active site of the ACE2 and their important interactions with various amino acid residues.

| Phytochemicals | B.E. (kcal/mole) | Important interactions at SGp-RBD docking site and catalytic sites (HIS345, HIS505 and ARG273) |
|---|---|---|
| Glycyrrhizinic acid | -9.5 | Hydrogen bond: ARG273, HIS374, TYR515, ASN394; Pi-Alkyl: PHE40, HIS40; Carbon hydrogen bond: GLU402; VDW: ARG393, TYR385, GLU402, ASP350, ALA348, TRP349, ASP382, HIS505, PHE504. |
| Maslinic acid | -8.5 | Hydrogen bond: PHE390, GLN388, ARG393, GLU37; Pi-Alkyl: VAL93, LYS26, PRO389; VDW: ASN33, ASP30, GLN96, THR92, ASN90. |
| Obacunone | -8.1 | Hydrogen bond: ARG273; Pi-Sigma: PHE504; Pi-Pi T-shaped: PHE504; Pi-Alkyl: TRP271, PHE504; VDW: PHE274, GLU145, HIS505, ASN149, LEU503, TYR127, ASN508, SER128. |
| Epoxyazadiradione | -8.0 | Alkyl/Pi-Alkyl: LYS26, PRO389; Pi-Sigma: HIS34; VDW: ASP30, ASN90, VAL93, GLN96, THR92, ASN33, GLU37. |
| Azadiradionolide | -8.0 | Hydrogen bond: HIS345, HIS401, ASN394; Alkyl/Pi-Alkyl: HIS373, ALA348, HIS374; VDW: PHE40, TRP349, ASP350, |

|  |  | THR347, GLU375, ARG514. |
|---|---|---|
| Ursolic acid | -7.4 | Hydrogen bond: LYS26, ASN90, ARG393; Pi-Alkyl: VAL93, PRO389, HIS34; VDW: ASP30, THR92, GLN96, ASN33, ALA387, GLU37, PHE390. |
| Gedunin | -7.3 | Hydrogen bond: HIS345; Alkyl/Pi-Alkyl: LUE370, PRO346, HIS374; Pi-Sigma: HIS374; Carbon hydrogen bond: PRO346; VDW: GLN442, ASP367, SER409, GLU406, GLU402, GLU375, THR371. |

## 4. Conclusions

In summary, we have screened 154 phytochemicals from limonoids and triterpenoids by molecular docking, *in silico* ADMET and drug-likeness prediction, and selected 15 phytochemicals to propose the potential hits against the five therapeutic protein targets (3CLpro, PLpro, RdRp, SpG-RBD and ACE2) of SARS-CoV-2. The phytochemicals 7-deacetyl-7-benzoylgedunin, glycyrrhizic acid, limonin and obacunone binds at the catalytic dyad of main protease 3CLpro. The phytochemicals obacunone, glycyrrhizic acid, ursolic acid and 7-deacetylgedunin binds at the catalytic triad of PLpro. Six phytochemicals glycyrrhizic acid, limonin, 7-deacetyl-7-benzoylgedunin, limonin glucoside, 7-deacetylgedunin and obacunone are found to bind at the active site of RdRp. The SGp-RBD site is important for the virion-host tropism, where the phytochemicals maslinic acid, glycyrrhizic acid, corosolic acid, 2-hydroxyseneganolide, oleanane and gedunin binds firmly with multiple non-covalent interactions. For the human ACE2 receptor, seven phytochemicals glycyrrhizinic acid, maslinic acid, obacunone, epoxyazadiradione, azadiradionolide, ursolic acid and gedunin were found binding at the catalytic site and/or the RBD site. Based on the dock score and reported medicinal properties, the combination of seven phytochemicals 7-deacetyl-7-benzoylgedunin, glycyrrhizic acid, limonin, obacunone, ursolic acid, corosolic acid and masilinic acid is sufficient to formulate an appropriate therapeutic approach to fight against SARS-CoV-2. The rich source of these phytochemicals are licorice, citrus, neem, holy basil and olives. Among the seven, the most important phytochemical is glycyrrhizic acid that binds at the active site of all the five protein targets of SARS-CoV-2. Overall, the computational predictions along with the reported pharmacological properties postulated that the limonoids and triterpenoids are potential against SARS-CoV-2 target proteins. The outcomes will be useful in formulating therapeutic strategies using the traditional medicines as well as the potential hits can be used for lead optimization for drug discovery for the treatment of COVID-19.

**Declaration of competing interest**

The authors declare that they have no known competing financial interests or personal relationships that could have appeared to influence the work reported in this paper.

**Acknowledgments**

Authors are thankful to Director, SVNIT for providing necessary research facilities and infrastructure.